\documentclass[%
 apl,
 jmp,%
 amsmath,amssymb,
 reprint,floatfix%
]{revtex4-1}
\usepackage{graphicx}
\usepackage{dcolumn}
\usepackage{bm}

\begin{document}

\preprint{AIP/123-QED}

\title{Highly Non-linear and Reliable Amorphous Silicon Based Back-to-Back Schottky Diode as Selector Device for Large Scale RRAM Arrays }

\author{Cheng-Chih Hsieh,$^{1}$ Yao-Feng Chang,$^{1}$ Ying-Chen Chen,$^{1}$ Heng-Lu Chang,$^{1}$ Sanjay.K.Banerjee$^{1}$}
\hspace*{-1cm}\affiliation{ 
Microelectronics Research Center, The University of Texas at Austin, 10100 Burnet Rd. Bldg. 160, Austin,Texas 78758, USA
}%
\author{Davood Shahrjerdi}%
\affiliation{%
Department of Electrical and Computer Engineering, New York University, 6 MetroTech Center, Brooklyn, New York 11201, USA
}%

\begin{abstract}
In this work we present silicon process compatible, stable and reliable ($>10^{8}$cycles), high non-linearity ratio at half-read voltage ($>5\times 10^{5}$), high speed ($<60ns$) low operating voltage ($<2V$) back-to-back Schottky diodes. Materials choice of electrode, thickness of semiconductor layer and doping level are investigated by numerical simulation, experiments and current-voltage equations to give a general design consideration when back-to-back Schottky diodes are used as selector device for Resistive Random Access Memory(RRAM) arrays.
\end{abstract}

\maketitle
\section*{}\vspace*{-0.95cm}
Resistive Random Access Memory (RRAM) is an emerging technology for future non-volatile memories (NVMs) because of fast write and read speed, low operating voltage, excellent scalability, very high density (4F$^2$) if using crossbar array architecture, with potential for three-dimensional integration. RRAM can be essential building block of neuromorphic copmuting.\cite{Strukov2008,Yang2008,Chang2016,Kuzum2013,Indiveri2013,Ebong2012,Pershin2009,Wei2008,Hsieh2015} However, RRAM crossbar arrays usually suffer from leakage current due to undesired sneak paths when selecting RRAM cells. As array size increases, more sneak paths can form in arrays thus it deteriorates read margin required to distinguish high and low resistance states. This needs to be addressed before moving on for any applications to large-scale RRAM arrays.\cite{Cappelletti2015,Chiquito2012} Non-linear (NL) devices in with RRAM cells which can provide very low current series through unselected cells while pass enough current to selected cells are highly desirable for realizing large scale RRAM arrays.This structure is usually called as one-selector-one-resistor (1S1R)\cite{Deng2013,Zhou2014,Huang2011,Lo2013,Srinivasan2012}. Compared to another structure, one-transistor-one-resistor (1T1R)\cite{Chen2013},  the benefit of 1S1R over 1T1R is that there is no area overhead for universal leakage-current-limiting devices, and making RRAM capable to build memory arrays with higher density than flash memory. Besides, the NL device must be Back-End-Of-Line (BEOL) compatible because most RRAM processes are currently integrated in BEOL. In this paper, we present highly NL, fast, robust, has low operating voltage, sufficient current density, good scalability, it uses a simple fabrication processes symmetric back-to-back Schottky diodes by metal-semiconductor-metal(MSM) structure. We also discuss impacts of material choices and device geometry on performance of MSM diode by numerical simulation, Schottky diode current characteristics and experimental results.

\begin{figure*}[htb]
\hspace*{-0.2cm}
\includegraphics[scale=0.64]{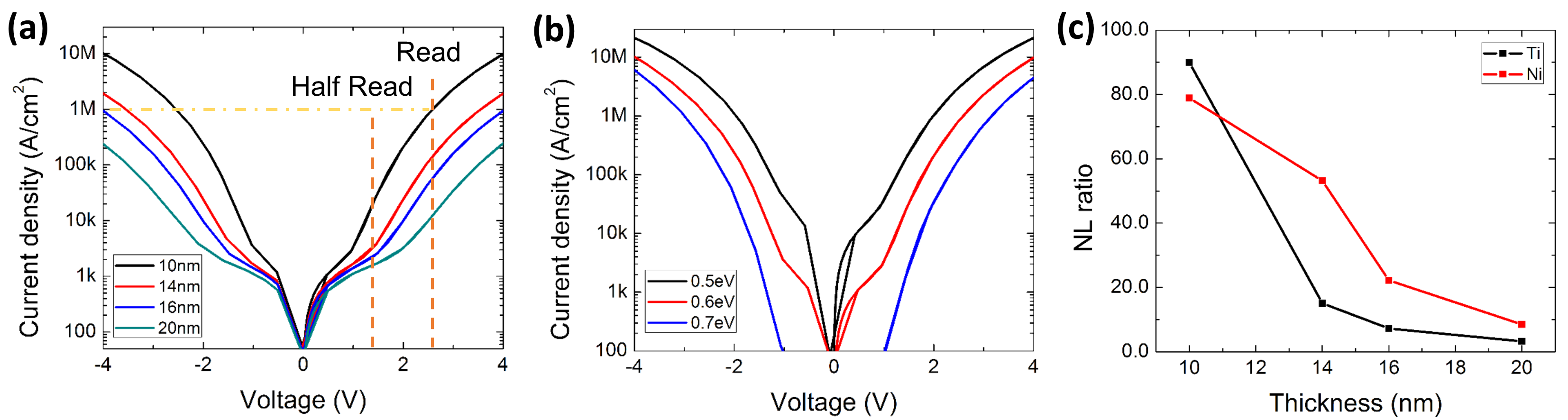}
\caption{(a) DC I-V characteristics from two-dimensional numerical simulation by Sentaurus from Synopsys Inc.. Two orange dash lines represent voltage values calculated for NL ratio. The choice of 1MA/cm$^{2}$ is based on matching current density of RRAM devices. (b) I-V characteristics on Schottky barrier height dependence. Note that non-linear step size of simulation caused I-V curves to show having a small hysteresis at low bias during positive polarity sweep, which is an artifact. (c) NL ratio extracted from Fig.1(a)(b). }
\end{figure*}

\section*{}\vspace*{-1.1cm}
We performed simulations of Sentaurus to evaluate the feasibility of MSM diode as selector device for future RRAM crossbar arrays.  We're interested in correlation between performance metrics of selector device and physical properties of MSM diode. In our simulation setup, we chose metal work function to match experimental values of Schottky barrier height values reported in literature which means we take Fermi-level pinning into account. We assume the dominant current transport mechanism in this ultra-thin MSM diode to be thermionic emission, recombination and tunneling. The doping concentration is assumed to be n-type doped (10$^{13}$cm$^{-3}$). DC characteristics in Fig.1(a) with different thickness for Schottky barrier height of 0.66eV, which is close to experimental values of titanium of silicon barrier\cite{Kuhn2008,Dimitriadis1995}. We see a thickness dependence on current density and NL ratio. The definition of NL ratio in this paper follows half-read scheme which is widely used in many memory systems. The NL ratio is defined as the current density ratio of 1 MA/cm$^{2}$ and the current at half of voltage where current density reaches 1 MA/cm$^{2}$. As thickness increases, NL ratio decreases. This is probably because the voltage region where current grows exponentially is delayed by higher resistance in the diode, and higher series resistance in thicker film also affects effective voltage drop in diode at higher current region. This makes I-V characteristics of thicker diodes deviate from ideal exponential curve at lower current compared to thinner diodes, thus NL ratio in thicker film is lower. The relation between Schottky barrier height and current density in Fig.1(b) agrees well with simple Schottky diode model. In our simulation, we found out the high doping concentration adversely affects NL ratio. This is intuitive since CMOS technology has been using highly doped source and drain to achieve ohmic contact, which contradict with our purpose to reach high NL ratio. We will discuss more details about current characteristics of MSM diode later, and the effect of Schottky barrier height in current characteristics of MSM diode is similar to simple Schottky diode. Higher Schottky barrier height results in lower current density near zero bias. One important design consideration of selector device is to have both high NL ratio and high current density at low voltage. Hence, based on the above simulation results, the desired MSM diode should have thin thickness of semiconductor layer (less than 14nm), low doping concentration and appropriate Schottky barrier height to give decent current density while keeping good NL ratio.
\section*{}\vspace*{-1.3cm}
After optimizing the design by numerical simulation, devices were fabricated on a n-type Si (111) substrate with 300 nm plasma-enhanced chemical vapor deposition (PECVD) grown silicon dioxide on the top as an isolation layer. 80nm bottom electrode (BE) was formed by electron beam evaporation at 273K. 10nm to 20nm semiconductor layer was deposited on the top of BE by PECVD at $250^\circ$C without \textit{ex-situ} annealing.The growth condition and parameters are referred to Moravej \textit{et al.}\cite{Moravej2004}, to aim to grow nanoscale hydrogenated amorphous silicon thin film. Then 80nm top electrode (TE) were formed by electron beam evaporation of titanium at 273K. Devices were patterned as crossbar with variant device perimeters from 300nm to 10$\mu$m by electron beam lithography. Electric measurements were taken by Agilent Semiconductor Parameter Analyzer B1500 and Lakeshore CRX-VF Probe Station. Pulse measurement was carried out by Agilent B1525 Pulse Generator Unit.  Devices were measured by applying voltage to the TE while the BE is grounded. The equivalent circuit diagram in Fig2.(a) illustrates series resistance in MSM diode. The design of the crossbar is targeted to minimize impacts on parasitic components when doing pulse measurements. It can be seen in Fig.2(b) metal lines are tapered to reduce parasitic capacitance. The overlap area of BE and TE defines the area of MSM diode.
\section*{}\vspace*{-1.2cm}
Fig.3(a) compares experimental DC I-V characteristics of Ti$-$aSi$-$Ti for different thickness of silicon layer. The effect of series resistance in 20nm amorphous Si MSM diode can be observed in the region highlighted by dashed circle. Ti$-$aSi$-$Ti has higher current density than Ni$-$aSi$-$Ni at the same thickness of amorphous silicon in Fig.3(b), which implies amorphous silicon is unintentionally doped with n-type impurities. For n-type silicon, nickle tends to pin closer to valence band of silicon while titanium is pinned at midgap. Thus we observe lower current density in Ni$-$aSi$-$Ni. NL ratio in Fig.3(c) agrees well with Fig.1(c), which also indicates series resistance plays an important role when designing high current density selector. The inset of Fig.3(c) shows linear scale plot of current density. It is obvious that there is asymmetry of current density between each polarity for titanium and nickel MSM diodes. Amorphous silicon and electrodes weren't deposited in the same instrument or same vacuum environment. So this causes inevitable interface difference between two Schottky diodes. To extract Schottky diode parameters from I-V characteristics of MSM diode, we need to start with simple Schottky diode current characteristics and combine with some assumptions to derive an insightful current equation for MSM diode device. Later we will point out when ideality factor larger than 1, MSM diode also demonstrates asymmetric I-V curve. Overall, experimental results are in good agreement with simulation prediction except of lower current density, probably because in simulation some materials-related parameters such as effective mass, recombination coefficients, and bandgap are not an very accurately known for amorphous silicon. 

\begin{figure*}[t]
\includegraphics[scale=0.7]{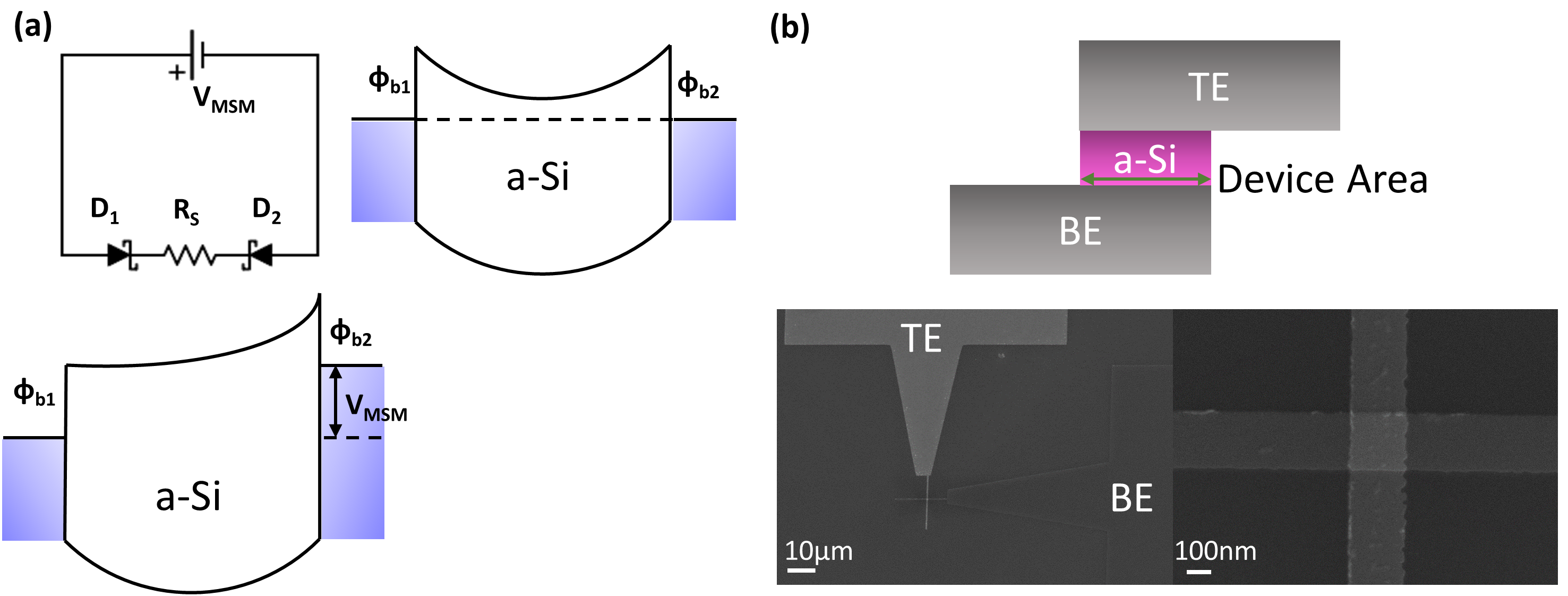}
\caption{(a) shows band diagrams at equilibrium and forward bias on D1. The equivalent circuit diagram was shown at forward bias on D1. Series resistance is labeled as R$_{S}$.$\phi_{b1}$ and $\phi_{b2}$ represents Schottky barrier height at D1 and D2. Fig 2(b) shows cartoon illustration of cross-sectional view of device. The green arrow indicates the effective area of device.}
\end{figure*}

\begin{figure*}[ht]
\includegraphics[scale=0.8]{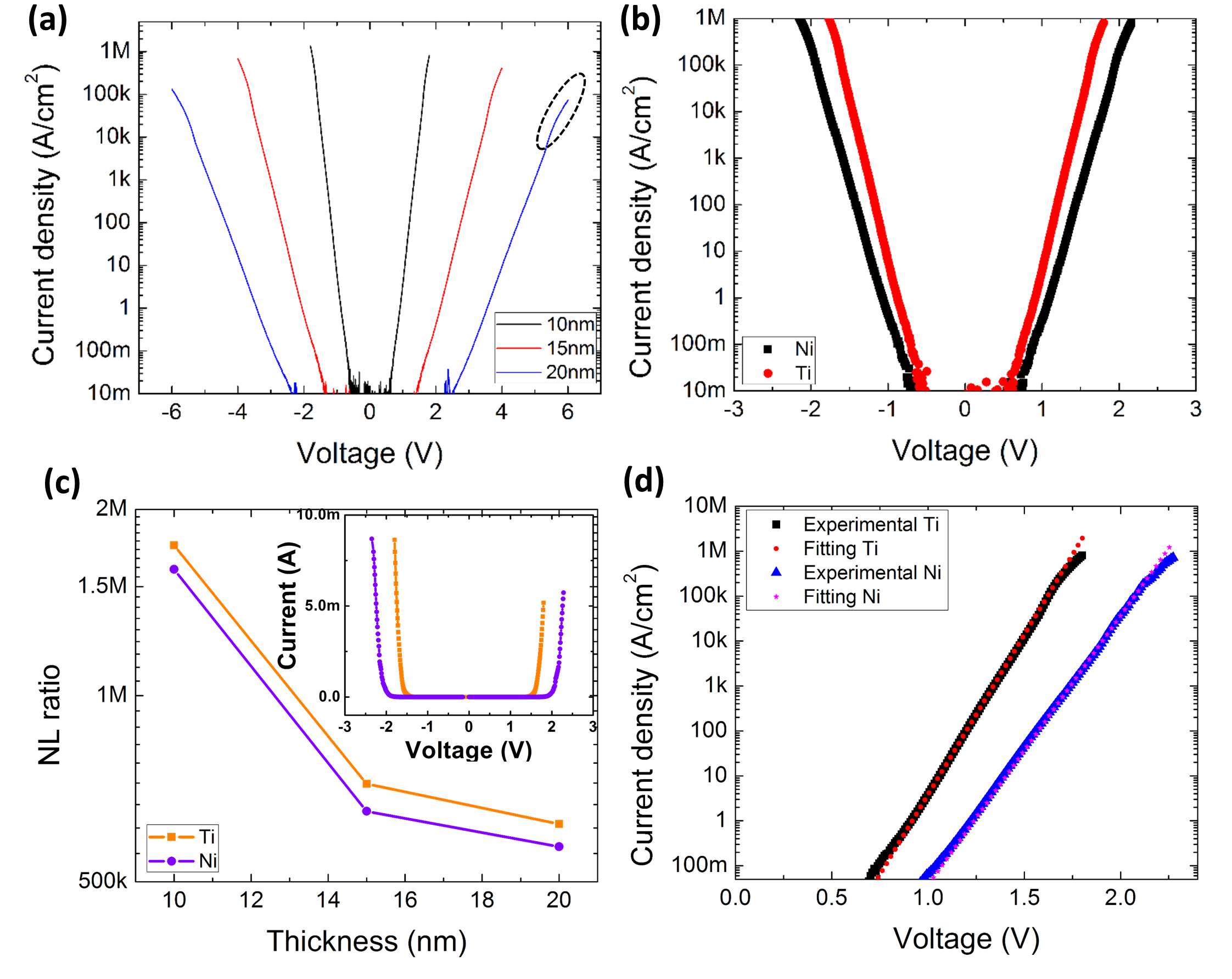}
\caption{(a) shows semilog plot of I-V characteristics of different thickness Ti$-$aSi$-$Ti diodes. The dashed circle is where series resistance effect becomes prominent. (b) shows I-V characteristics of Ti$-$aSi$-$Ti and Ni$-$aSi$-$Ni with 10nm amorphous silicon.(c) compares NL ratio between Ti$-$aSi$-$Ti and Ni$-$aSi$-$Ni at different thickness. The inset of (c) is linear scale I-V characteristics of (b). (d) shows I-V curve fitting by using $(7)$. $(7)$ fits well at intermediate bias where series resistance effect was low and thermionic emission is dominant.}
\end{figure*}

\begin{figure*}[htb]
\includegraphics[scale=0.72]{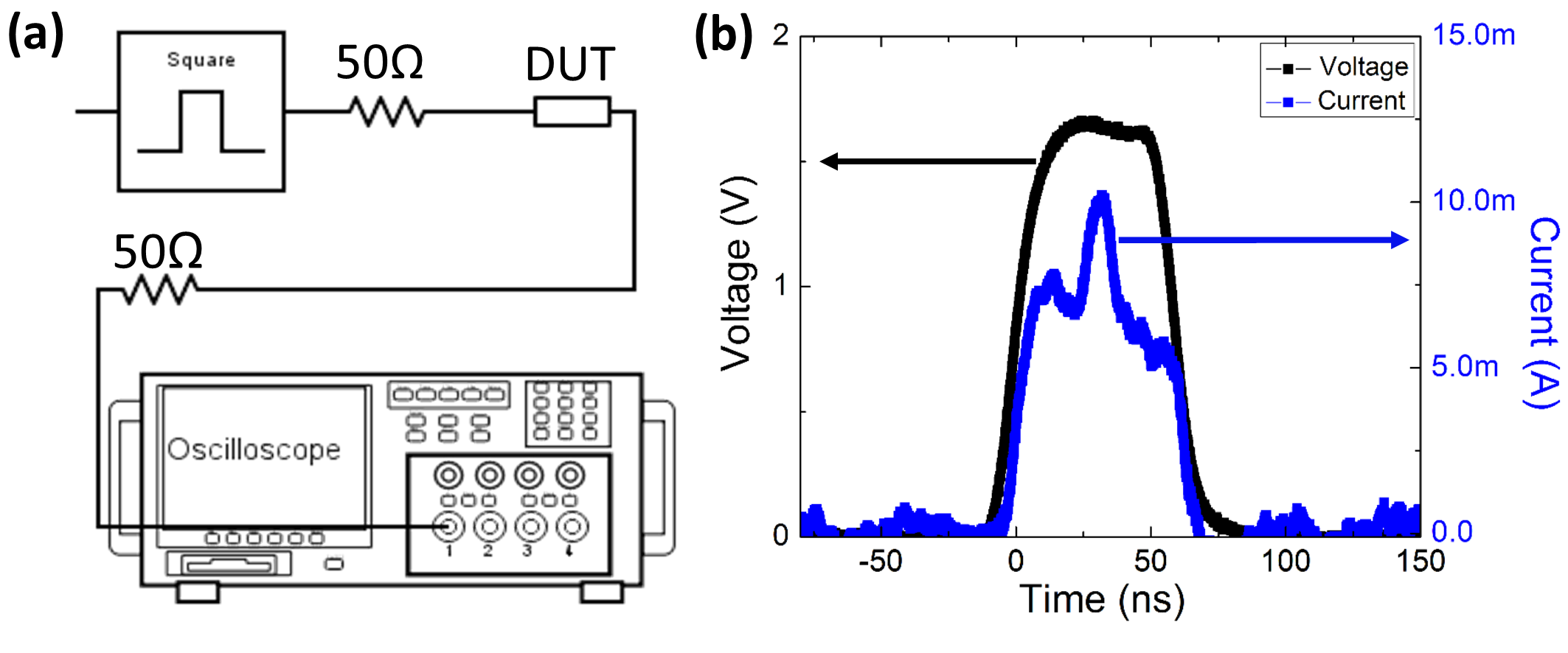}
\caption{(a) Pulse measurement setup and impedance at each stage.The pulse generator has an output impedence of 50 ohm, and input channel impedance of oscilloscope is 50 ohm as well. Fig4.(b) shows the transient current response of DUT. black curve and arrow is input pulse generated by source and blue curve and arrow is current response of MSM diode. The device dimension is 300nm by 300nm crossbar.}
\end{figure*}

\begin{figure*}[ht]
\includegraphics[scale=0.68]{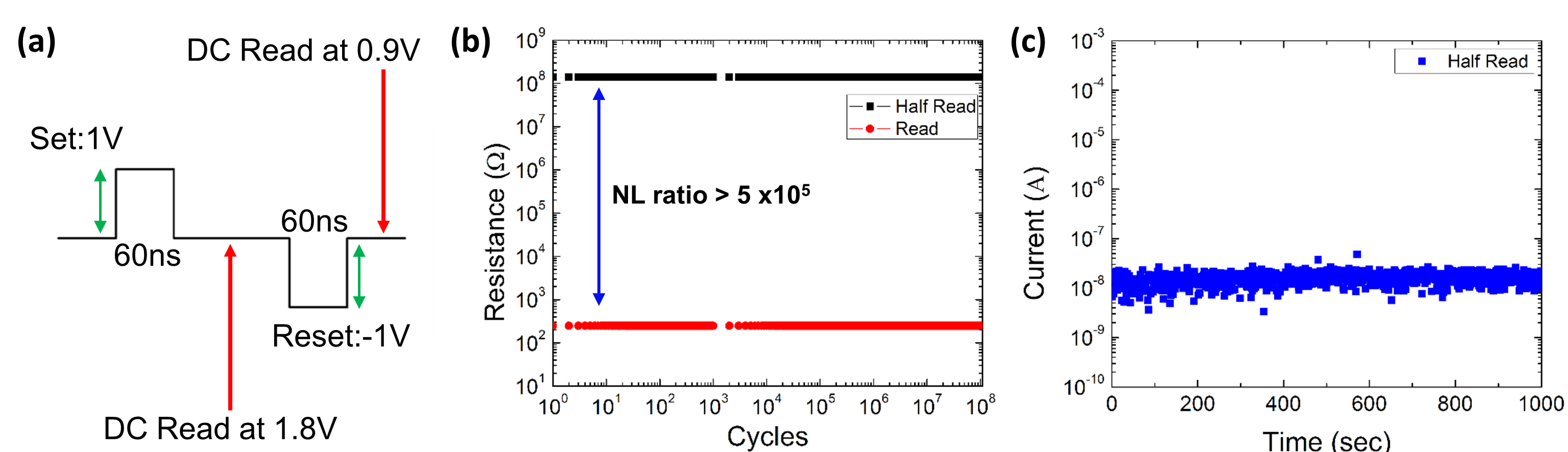}
\caption{(a) Test scheme of cycle test. Fig.5(b) shows cycle test of 10$^{8}$ cycles. Fig.5(c) shows DC stress test at 0.9V for 1000 seconds.}
\end{figure*}
\section*{}\vspace*{-1.3cm}
We want to understand current-voltage characteristics of MSM diode so that we can extract fundamental parameters such as Schottky barrier height and ideality factor by fitting experimental I-V curves. The band diagram in Fig.2(a) indicates that MSM diode can be considered to be back-to-back Schottky diodes. The current-voltage equation of a simple Schottky diode is in the following form: 
\begin{equation}
J=J_S\exp{(\frac{qV}{\eta kT})}
\end{equation}
J$_s$ is reverse saturation current of Schottky diode:
\begin{equation}
J_s=A*T^2\exp{(\frac{q\phi_{b}^{0}}{kT})}
\end{equation}
where $A$ is Richardson constant, $T$ is temperature and $k$ is Boltzmann constant. The voltage drops in two Schottky diodes are $V_{1}$ and $V_{2}$ and voltage across MSM diode is $V_{MSM}= V_{1}+V_{2}$. We can also write $J_{1}=-J_{2}$ from current continuity. Based on these two conditions we can write current-voltage equation as in Nouchi\textit{et al.} and Nagano\textit{et al.}\cite{Nouchi2014,Nagano2007}:
\begin{equation}
J_{MSM}=\frac{2J_{S1}J_{S2}\sinh(\frac{qV_{MSM}}{2\eta kT})}{J_{S1}\exp{(\frac{qV_{MSM}}{2\eta kT})}+J_{S2}\exp{(\frac{-qV_{MSM}}{2\eta kT})}}
\end{equation}
The equation above can be further simplified if Schottky barrier height of diode 1 and diode 2 are symmetric:
\begin{equation}
J_{MSM}=J_{S}\tanh(\frac{qV_{MSM}}{2\eta kT})
\end{equation}
However, the current obtained from $(4)$ reaches saturation at very low bias; this is contradicting to reported experimental results\cite{Bozyigit2015,Berger1996} and our experimental data. Note that $(4)$ doesn't consider image-force lowering. Image-force lowering is based on image charges in metal layer induced by charges in semiconductor layer near metal-semiconductor interface. Image charges establish an electric field along metal-semiconductor interface and Schottky barrier height becomes voltage dependent:
\begin{equation}
\phi_{b1}(V_{1})=\phi_{b1}^{0}+qV_{1}(1-\frac{1}{\eta})
\end{equation}
\begin{equation}
\phi_{b2}(V_{2})=\phi_{b2}^{0}+qV_{2}(1-\frac{1}{\eta})
\end{equation}
where $\phi_{b}^{0}$ is Schottky barrier height under zero bias and $\eta$ is ideality factor. Assuming voltage drop in MSM diode is mostly on diode 2 since diode 1 is forward bias in Fig.2(a), we have $V_{MSM}\approx V_{2}$. We also assume $(5)$ equals to $(6)$ because of symmetric MSM structure, then we can rewrite $(1)$:
\begin{equation}
J=J_{S}\sinh(\frac{qV_{MSM}}{2kT})\exp{(\frac{qV_{MSM}}{2kT})}\exp{(\frac{-qV_{MSM}}{\eta kT})}
\end{equation}
With Eq.$(7)$ we can evaluate Schottky barrier height and ideality factor in MSM diodes and discuss the impacts on performance matrices of selector device.Fig.3(d) shows fitting by $(7)$ on 10nm Ti$-$aSi$-$Ti and Ni$-$aSi$-$Ni MSM diode. The Schottky barrier height between titanium and amorphous silicon extracted by curve fitting with Eq.$(7)$ is $\approx$ 0.79$eV$, and the Schottky barrier height between nickle and amorphous silicon is about 0.86$eV$.Note that the bandgap of hydrogenated amorphous silicon is around 1.6-1.8$eV$.Titanium is usually pinned at midgap at silicon interface and nickel is usually pinned closer to valence band of silicon, so the barrier height values extracted from Fig3.(d) are what we expect. The ideality factor extracted from Fig.3(d) are 1.18 and 1.36 for titanium and nickel respectively. Eq.$(7)$ implies that non-ideal ideality factor can be  attributed to unequal current density in MSM diode. The magnitude of current density difference between each polarity is sensitive to ideality factor. As mentioned in the previous section, two Schottky diodes might have slightly different interfaces with silicon. This might cause slightly different Schottky barrier height and thus contribute to asymmetric I-V characteristics. Although there is asymmetric current density for the same voltage when applying different polarity, the ratio of current density asymmetry is less than one order, as seen in Fig.3(c). The voltage difference to reach 1MA/cm$^{2}$ is only 0.1V for different voltage polarity for both Ti$-$aSi$-$Ti and Ni$-$aSi$-$Ni MSM diode. This will not impact the operation of selector device, but the effect of non-ideal MSM diode and interface properties should be taken into account when applying the selector device in RRAM arrays.
\section*{}\vspace*{-1.2cm}
We performed transient analysis to test the speed of Ti$-$aSi$-$Ti device. The test setup is shown in Fig4.(a). We did impedance matching for the source to ensure the actual voltage drop in device under test (DUT) was correct and we normalized the current scale measured at oscilloscope because input channel impedance was 50 ohm but the impedance through DUT was a few thousands ohm. It can be seen that in Fig4.(b) MSM diode can response to 60ns pulse without any notable delay or distortion of signal. The current overshoot was only 25\% higher than the mean current level during pulses. The result of transient anaylsis is reasonable because the dominant current transport in MSM diode is thermionic emission. This makes MSM diode favorable over pn junction diode which is based on minority carrier injection.
\section*{}\vspace*{-1.2cm}
MSM diode also shows excellent reliability as potential selector device. Fig.5(a) illustrates cycle test scheme of MSM diode. This is very similar to the test scheme of RRAM device since selector devices need to be integrated with RRAM cells. After each set or reset pulse, DC read at read voltage or half read voltage was performed to record the value of resistance. Fig.5(b) shows MSM diodes survived after 10$^{8}$ cycles without any notable degradation. Besides, MSM diodes demonstrated very good DC stress test, as seen in Fig.5(c) at half-read voltage over 10$^{3}$ seconds. These test results validate that our MSM diode is very robust and a suitable candidate of selector device.
\section*{}\vspace*{-1.2cm}
In summary, a fast, reliable, high NL ratio, low operating voltage with large current density MSM diode based selector device is presented in this paper. The current-voltage characteristics of symmetric MSM diode is also derived to extract Schottky barrier height between metal and amorphous silicon and ideality factor. The current density asymmetry at a given voltage between each polarity is due to non-ideal MSM diode and different interface properties at each diode. The performance of MSM diodes could be improved if we use a narrower bandgap semiconductor or choose lower Schottky barrier height metal but one needs to consider fabrication feasibility and reliability of selector device. Here, we present an idea of designing MSM diode to meet requirements for selector device.
\section*{}\vspace*{-1.2cm}
This work was supported by National Science Foundation (NSF), National Nanotechnologies Coordinated Infrastructure (NNCI), and Nanomanufacturing
Systems for Mobile Computing and Mobile Energy Technologies (NASCENT).
\begin{align*}
\end{align*}
\bibliography{APL_MSM_diode.bib}

\begin{thebibliography}{24}%
\makeatletter
\providecommand \@ifxundefined [1]{%
 \@ifx{#1\undefined}
}%
\providecommand \@ifnum [1]{%
 \ifnum #1\expandafter \@firstoftwo
 \else \expandafter \@secondoftwo
 \fi
}%
\providecommand \@ifx [1]{%
 \ifx #1\expandafter \@firstoftwo
 \else \expandafter \@secondoftwo
 \fi
}%
\providecommand \natexlab [1]{#1}%
\providecommand \enquote  [1]{``#1''}%
\providecommand \bibnamefont  [1]{#1}%
\providecommand \bibfnamefont [1]{#1}%
\providecommand \citenamefont [1]{#1}%
\providecommand \href@noop [0]{\@secondoftwo}%
\providecommand \href [0]{\begingroup \@sanitize@url \@href}%
\providecommand \@href[1]{\@@startlink{#1}\@@href}%
\providecommand \@@href[1]{\endgroup#1\@@endlink}%
\providecommand \@sanitize@url [0]{\catcode `\\12\catcode `\$12\catcode
  `\&12\catcode `\#12\catcode `\^12\catcode `\_12\catcode `\%12\relax}%
\providecommand \@@startlink[1]{}%
\providecommand \@@endlink[0]{}%
\providecommand \url  [0]{\begingroup\@sanitize@url \@url }%
\providecommand \@url [1]{\endgroup\@href {#1}{\urlprefix }}%
\providecommand \urlprefix  [0]{URL }%
\providecommand \Eprint [0]{\href }%
\providecommand \doibase [0]{http://dx.doi.org/}%
\providecommand \selectlanguage [0]{\@gobble}%
\providecommand \bibinfo  [0]{\@secondoftwo}%
\providecommand \bibfield  [0]{\@secondoftwo}%
\providecommand \translation [1]{[#1]}%
\providecommand \BibitemOpen [0]{}%
\providecommand \bibitemStop [0]{}%
\providecommand \bibitemNoStop [0]{.\EOS\space}%
\providecommand \EOS [0]{\spacefactor3000\relax}%
\providecommand \BibitemShut  [1]{\csname bibitem#1\endcsname}%
\let\auto@bib@innerbib\@empty
\bibitem [{\citenamefont {Strukov}\ \emph {et~al.}(2008)\citenamefont
  {Strukov}, \citenamefont {Snider}, \citenamefont {Stewart},\ and\
  \citenamefont {Williams}}]{Strukov2008}%
  \BibitemOpen
  \bibfield  {author} {\bibinfo {author} {\bibfnamefont {D.~B.}\ \bibnamefont
  {Strukov}}, \bibinfo {author} {\bibfnamefont {G.~S.}\ \bibnamefont {Snider}},
  \bibinfo {author} {\bibfnamefont {D.~R.}\ \bibnamefont {Stewart}}, \ and\
  \bibinfo {author} {\bibfnamefont {R.~S.}\ \bibnamefont {Williams}},\ }\href
  {\doibase 10.1038/nature06932} {\bibfield  {journal} {\bibinfo  {journal}
  {Nature}\ }\textbf {\bibinfo {volume} {453}},\ \bibinfo {pages} {80}
  (\bibinfo {year} {2008})}\BibitemShut {NoStop}%
\bibitem [{\citenamefont {Yang}\ \emph {et~al.}(2008)\citenamefont {Yang},
  \citenamefont {Pickett}, \citenamefont {Li}, \citenamefont {Ohlberg},
  \citenamefont {Stewart},\ and\ \citenamefont {Williams}}]{Yang2008}%
  \BibitemOpen
  \bibfield  {author} {\bibinfo {author} {\bibfnamefont {J.~J.}\ \bibnamefont
  {Yang}}, \bibinfo {author} {\bibfnamefont {M.~D.}\ \bibnamefont {Pickett}},
  \bibinfo {author} {\bibfnamefont {X.}~\bibnamefont {Li}}, \bibinfo {author}
  {\bibfnamefont {D.~A.~A.}\ \bibnamefont {Ohlberg}}, \bibinfo {author}
  {\bibfnamefont {D.~R.}\ \bibnamefont {Stewart}}, \ and\ \bibinfo {author}
  {\bibfnamefont {R.~S.}\ \bibnamefont {Williams}},\ }\href {\doibase
  10.1038/nnano.2008.160} {\bibfield  {journal} {\bibinfo  {journal} {Nat.
  Nanotechnol.}\ }\textbf {\bibinfo {volume} {3}},\ \bibinfo {pages} {429}
  (\bibinfo {year} {2008})}\BibitemShut {NoStop}%
\bibitem [{\citenamefont {Chang}\ \emph {et~al.}(2016)\citenamefont {Chang},
  \citenamefont {Fowler}, \citenamefont {Chen}, \citenamefont {Zhou},
  \citenamefont {Pan}, \citenamefont {Chang},\ and\ \citenamefont
  {Lee}}]{Chang2016}%
  \BibitemOpen
  \bibfield  {author} {\bibinfo {author} {\bibfnamefont {Y.-F.}\ \bibnamefont
  {Chang}}, \bibinfo {author} {\bibfnamefont {B.}~\bibnamefont {Fowler}},
  \bibinfo {author} {\bibfnamefont {Y.-C.}\ \bibnamefont {Chen}}, \bibinfo
  {author} {\bibfnamefont {F.}~\bibnamefont {Zhou}}, \bibinfo {author}
  {\bibfnamefont {C.-H.}\ \bibnamefont {Pan}}, \bibinfo {author} {\bibfnamefont
  {T.-C.}\ \bibnamefont {Chang}}, \ and\ \bibinfo {author} {\bibfnamefont
  {J.~C.}\ \bibnamefont {Lee}},\ }\href {\doibase 10.1038/srep21268} {\bibfield
   {journal} {\bibinfo  {journal} {Sci. Rep.}\ }\textbf {\bibinfo {volume}
  {6}},\ \bibinfo {pages} {21268} (\bibinfo {year} {2016})}\BibitemShut
  {NoStop}%
\bibitem [{\citenamefont {Kuzum}\ \emph {et~al.}(2013)\citenamefont {Kuzum},
  \citenamefont {Yu},\ and\ \citenamefont {Wong}}]{Kuzum2013}%
  \BibitemOpen
  \bibfield  {author} {\bibinfo {author} {\bibfnamefont {D.}~\bibnamefont
  {Kuzum}}, \bibinfo {author} {\bibfnamefont {S.}~\bibnamefont {Yu}}, \ and\
  \bibinfo {author} {\bibfnamefont {H.-S.~P.}\ \bibnamefont {Wong}},\ }\href
  {\doibase 10.1088/0957-4484/24/38/382001} {\bibfield  {journal} {\bibinfo
  {journal} {Nanotechnology}\ }\textbf {\bibinfo {volume} {24}},\ \bibinfo
  {pages} {382001} (\bibinfo {year} {2013})}\BibitemShut {NoStop}%
\bibitem [{\citenamefont {Indiveri}\ \emph {et~al.}(2013)\citenamefont
  {Indiveri}, \citenamefont {Linares-Barranco}, \citenamefont {Legenstein},
  \citenamefont {Deligeorgis},\ and\ \citenamefont
  {Prodromakis}}]{Indiveri2013}%
  \BibitemOpen
  \bibfield  {author} {\bibinfo {author} {\bibfnamefont {G.}~\bibnamefont
  {Indiveri}}, \bibinfo {author} {\bibfnamefont {B.}~\bibnamefont
  {Linares-Barranco}}, \bibinfo {author} {\bibfnamefont {R.}~\bibnamefont
  {Legenstein}}, \bibinfo {author} {\bibfnamefont {G.}~\bibnamefont
  {Deligeorgis}}, \ and\ \bibinfo {author} {\bibfnamefont {T.}~\bibnamefont
  {Prodromakis}},\ }\href {\doibase 10.1088/0957-4484/24/38/384010} {\bibfield
  {journal} {\bibinfo  {journal} {Nanotechnology}\ }\textbf {\bibinfo {volume}
  {24}},\ \bibinfo {pages} {384010} (\bibinfo {year} {2013})}\BibitemShut
  {NoStop}%
\bibitem [{\citenamefont {Ebong}\ and\ \citenamefont
  {Mazumder}(2012)}]{Ebong2012}%
  \BibitemOpen
  \bibfield  {author} {\bibinfo {author} {\bibfnamefont {I.~E.}\ \bibnamefont
  {Ebong}}\ and\ \bibinfo {author} {\bibfnamefont {P.}~\bibnamefont
  {Mazumder}},\ }\href {\doibase 10.1109/JPROC.2011.2173089} {\bibfield
  {journal} {\bibinfo  {journal} {Proc. IEEE}\ }\textbf {\bibinfo {volume}
  {100}},\ \bibinfo {pages} {2050} (\bibinfo {year} {2012})}\BibitemShut
  {NoStop}%
\bibitem [{\citenamefont {Pershin}\ \emph {et~al.}(2009)\citenamefont
  {Pershin}, \citenamefont {{La Fontaine}},\ and\ \citenamefont {{Di
  Ventra}}}]{Pershin2009}%
  \BibitemOpen
  \bibfield  {author} {\bibinfo {author} {\bibfnamefont {Y.~V.}\ \bibnamefont
  {Pershin}}, \bibinfo {author} {\bibfnamefont {S.}~\bibnamefont {{La
  Fontaine}}}, \ and\ \bibinfo {author} {\bibfnamefont {M.}~\bibnamefont {{Di
  Ventra}}},\ }\href {\doibase 10.1103/PhysRevE.80.021926} {\bibfield
  {journal} {\bibinfo  {journal} {Phys. Rev. E - Stat. Nonlinear, Soft Matter
  Phys.}\ }\textbf {\bibinfo {volume} {80}},\ \bibinfo {pages} {1} (\bibinfo
  {year} {2009})},\ \Eprint {http://arxiv.org/abs/0810.4179} {arXiv:0810.4179}
  \BibitemShut {NoStop}%
\bibitem [{\citenamefont {Wei}\ \emph {et~al.}(2008)\citenamefont {Wei},
  \citenamefont {Kanzawa}, \citenamefont {Arita}, \citenamefont {Katoh},
  \citenamefont {Kawai}, \citenamefont {Muraoka}, \citenamefont {Mitani},
  \citenamefont {Fujii}, \citenamefont {Katayama}, \citenamefont {Iijima},
  \citenamefont {Mikawa}, \citenamefont {Ninomiya}, \citenamefont {Miyanaga},
  \citenamefont {Kawashima}, \citenamefont {Tsuji}, \citenamefont {Himeno},
  \citenamefont {Okada}, \citenamefont {Azuma}, \citenamefont {Shimakawa},
  \citenamefont {Sugaya}, \citenamefont {Takagi}, \citenamefont {Yasuhara},
  \citenamefont {Horiba}, \citenamefont {Kumigashira},\ and\ \citenamefont
  {Oshima}}]{Wei2008}%
  \BibitemOpen
  \bibfield  {author} {\bibinfo {author} {\bibfnamefont {Z.}~\bibnamefont
  {Wei}}, \bibinfo {author} {\bibfnamefont {Y.}~\bibnamefont {Kanzawa}},
  \bibinfo {author} {\bibfnamefont {K.}~\bibnamefont {Arita}}, \bibinfo
  {author} {\bibfnamefont {Y.}~\bibnamefont {Katoh}}, \bibinfo {author}
  {\bibfnamefont {K.}~\bibnamefont {Kawai}}, \bibinfo {author} {\bibfnamefont
  {S.}~\bibnamefont {Muraoka}}, \bibinfo {author} {\bibfnamefont
  {S.}~\bibnamefont {Mitani}}, \bibinfo {author} {\bibfnamefont
  {S.}~\bibnamefont {Fujii}}, \bibinfo {author} {\bibfnamefont
  {K.}~\bibnamefont {Katayama}}, \bibinfo {author} {\bibfnamefont
  {M.}~\bibnamefont {Iijima}}, \bibinfo {author} {\bibfnamefont
  {T.}~\bibnamefont {Mikawa}}, \bibinfo {author} {\bibfnamefont
  {T.}~\bibnamefont {Ninomiya}}, \bibinfo {author} {\bibfnamefont
  {R.}~\bibnamefont {Miyanaga}}, \bibinfo {author} {\bibfnamefont
  {Y.}~\bibnamefont {Kawashima}}, \bibinfo {author} {\bibfnamefont
  {K.}~\bibnamefont {Tsuji}}, \bibinfo {author} {\bibfnamefont
  {A.}~\bibnamefont {Himeno}}, \bibinfo {author} {\bibfnamefont
  {T.}~\bibnamefont {Okada}}, \bibinfo {author} {\bibfnamefont
  {R.}~\bibnamefont {Azuma}}, \bibinfo {author} {\bibfnamefont
  {K.}~\bibnamefont {Shimakawa}}, \bibinfo {author} {\bibfnamefont
  {H.}~\bibnamefont {Sugaya}}, \bibinfo {author} {\bibfnamefont
  {T.}~\bibnamefont {Takagi}}, \bibinfo {author} {\bibfnamefont
  {R.}~\bibnamefont {Yasuhara}}, \bibinfo {author} {\bibfnamefont
  {K.}~\bibnamefont {Horiba}}, \bibinfo {author} {\bibfnamefont
  {H.}~\bibnamefont {Kumigashira}}, \ and\ \bibinfo {author} {\bibfnamefont
  {M.}~\bibnamefont {Oshima}},\ }\href {\doibase 10.1109/IEDM.2008.4796676}
  {\bibfield  {journal} {\bibinfo  {journal} {2008 IEEE Int. Electron Devices
  Meet.}\ ,\ \bibinfo {pages} {1}} (\bibinfo {year} {2008})}\BibitemShut
  {NoStop}%
\bibitem [{\citenamefont {Hsieh}\ \emph {et~al.}(2015)\citenamefont {Hsieh},
  \citenamefont {Roy}, \citenamefont {Rai}, \citenamefont {Chang},\ and\
  \citenamefont {Banerjee}}]{Hsieh2015}%
  \BibitemOpen
  \bibfield  {author} {\bibinfo {author} {\bibfnamefont {C.-C.}\ \bibnamefont
  {Hsieh}}, \bibinfo {author} {\bibfnamefont {A.}~\bibnamefont {Roy}}, \bibinfo
  {author} {\bibfnamefont {A.}~\bibnamefont {Rai}}, \bibinfo {author}
  {\bibfnamefont {Y.-F.}\ \bibnamefont {Chang}}, \ and\ \bibinfo {author}
  {\bibfnamefont {S.~K.}\ \bibnamefont {Banerjee}},\ }\href {\doibase
  10.1063/1.4919442} {\bibfield  {journal} {\bibinfo  {journal} {Appl. Phys.
  Lett.}\ }\textbf {\bibinfo {volume} {106}},\ \bibinfo {pages} {173108}
  (\bibinfo {year} {2015})}\BibitemShut {NoStop}%
\bibitem [{\citenamefont {Cappelletti}(2015)}]{Cappelletti2015}%
  \BibitemOpen
  \bibfield  {author} {\bibinfo {author} {\bibfnamefont {P.}~\bibnamefont
  {Cappelletti}},\ }\href@noop {} {\bibfield  {journal} {\bibinfo  {journal}
  {2015 IEEE Int. Electron Devices Meet.}\ ,\ \bibinfo {pages} {241}} (\bibinfo
  {year} {2015})}\BibitemShut {NoStop}%
\bibitem [{\citenamefont {Chiquito}\ \emph {et~al.}(2012)\citenamefont
  {Chiquito}, \citenamefont {Amorim}, \citenamefont {Berengue}, \citenamefont
  {Araujo}, \citenamefont {Bernardo},\ and\ \citenamefont
  {Leite}}]{Chiquito2012}%
  \BibitemOpen
  \bibfield  {author} {\bibinfo {author} {\bibfnamefont {A.~J.}\ \bibnamefont
  {Chiquito}}, \bibinfo {author} {\bibfnamefont {C.~a.}\ \bibnamefont
  {Amorim}}, \bibinfo {author} {\bibfnamefont {O.~M.}\ \bibnamefont
  {Berengue}}, \bibinfo {author} {\bibfnamefont {L.~S.}\ \bibnamefont
  {Araujo}}, \bibinfo {author} {\bibfnamefont {E.~P.}\ \bibnamefont
  {Bernardo}}, \ and\ \bibinfo {author} {\bibfnamefont {E.~R.}\ \bibnamefont
  {Leite}},\ }\href {\doibase 10.1088/0953-8984/24/22/225303} {\bibfield
  {journal} {\bibinfo  {journal} {J. Phys. Condens. Matter}\ }\textbf {\bibinfo
  {volume} {24}},\ \bibinfo {pages} {225303} (\bibinfo {year}
  {2012})}\BibitemShut {NoStop}%
\bibitem [{\citenamefont {Deng}\ \emph {et~al.}(2013)\citenamefont {Deng},
  \citenamefont {Huang}, \citenamefont {Chen}, \citenamefont {Yang},
  \citenamefont {Gao}, \citenamefont {Wang}, \citenamefont {Zeng},
  \citenamefont {Du}, \citenamefont {Kang},\ and\ \citenamefont
  {Liu}}]{Deng2013}%
  \BibitemOpen
  \bibfield  {author} {\bibinfo {author} {\bibfnamefont {Y.}~\bibnamefont
  {Deng}}, \bibinfo {author} {\bibfnamefont {P.}~\bibnamefont {Huang}},
  \bibinfo {author} {\bibfnamefont {B.}~\bibnamefont {Chen}}, \bibinfo {author}
  {\bibfnamefont {X.}~\bibnamefont {Yang}}, \bibinfo {author} {\bibfnamefont
  {B.}~\bibnamefont {Gao}}, \bibinfo {author} {\bibfnamefont {J.}~\bibnamefont
  {Wang}}, \bibinfo {author} {\bibfnamefont {L.}~\bibnamefont {Zeng}}, \bibinfo
  {author} {\bibfnamefont {G.}~\bibnamefont {Du}}, \bibinfo {author}
  {\bibfnamefont {J.}~\bibnamefont {Kang}}, \ and\ \bibinfo {author}
  {\bibfnamefont {X.}~\bibnamefont {Liu}},\ }\href {\doibase
  10.1109/TED.2012.2231683} {\bibfield  {journal} {\bibinfo  {journal} {IEEE
  Trans. Electron Devices}\ }\textbf {\bibinfo {volume} {60}},\ \bibinfo
  {pages} {719} (\bibinfo {year} {2013})}\BibitemShut {NoStop}%
\bibitem [{\citenamefont {Zhou}\ \emph {et~al.}(2014)\citenamefont {Zhou},
  \citenamefont {Kim},\ and\ \citenamefont {Lu}}]{Zhou2014}%
  \BibitemOpen
  \bibfield  {author} {\bibinfo {author} {\bibfnamefont {J.}~\bibnamefont
  {Zhou}}, \bibinfo {author} {\bibfnamefont {K.-H.}\ \bibnamefont {Kim}}, \
  and\ \bibinfo {author} {\bibfnamefont {W.}~\bibnamefont {Lu}},\ }\href
  {\doibase 10.1109/TED.2014.2310200} {\bibfield  {journal} {\bibinfo
  {journal} {Electron Devices, IEEE Trans.}\ }\textbf {\bibinfo {volume}
  {61}},\ \bibinfo {pages} {1369} (\bibinfo {year} {2014})}\BibitemShut
  {NoStop}%
\bibitem [{\citenamefont {Huang}\ \emph {et~al.}(2011)\citenamefont {Huang},
  \citenamefont {Tseng}, \citenamefont {Luo}, \citenamefont {Hsu},\ and\
  \citenamefont {Hou}}]{Huang2011}%
  \BibitemOpen
  \bibfield  {author} {\bibinfo {author} {\bibfnamefont {J.~J.}\ \bibnamefont
  {Huang}}, \bibinfo {author} {\bibfnamefont {Y.~M.}\ \bibnamefont {Tseng}},
  \bibinfo {author} {\bibfnamefont {W.~C.}\ \bibnamefont {Luo}}, \bibinfo
  {author} {\bibfnamefont {C.~W.}\ \bibnamefont {Hsu}}, \ and\ \bibinfo
  {author} {\bibfnamefont {T.~H.}\ \bibnamefont {Hou}},\ }in\ \href {\doibase
  10.1109/IEDM.2011.6131653} {\emph {\bibinfo {booktitle} {Tech. Dig. - Int.
  Electron Devices Meet. IEDM}}}\ (\bibinfo {year} {2011})\ pp.\ \bibinfo
  {pages} {733--736}\BibitemShut {NoStop}%
\bibitem [{\citenamefont {Lo}\ \emph {et~al.}(2013)\citenamefont {Lo},
  \citenamefont {Chen}, \citenamefont {Huang},\ and\ \citenamefont
  {Hou}}]{Lo2013}%
  \BibitemOpen
  \bibfield  {author} {\bibinfo {author} {\bibfnamefont {C.~L.}\ \bibnamefont
  {Lo}}, \bibinfo {author} {\bibfnamefont {M.~C.}\ \bibnamefont {Chen}},
  \bibinfo {author} {\bibfnamefont {J.~J.}\ \bibnamefont {Huang}}, \ and\
  \bibinfo {author} {\bibfnamefont {T.~H.}\ \bibnamefont {Hou}},\ }in\ \href
  {\doibase 10.1109/VLSI-TSA.2013.6545588} {\emph {\bibinfo {booktitle} {2013
  Int. Symp. VLSI Technol. Syst. Appl. VLSI-TSA 2013}}}\ (\bibinfo {year}
  {2013})\ pp.\ \bibinfo {pages} {0--1}\BibitemShut {NoStop}%
\bibitem [{\citenamefont {Srinivasan}\ \emph {et~al.}(2012)\citenamefont
  {Srinivasan}, \citenamefont {Chopra}, \citenamefont {Karkare}, \citenamefont
  {Bafna}, \citenamefont {Lashkare}, \citenamefont {Kumbhare}, \citenamefont
  {Kim}, \citenamefont {Srinivasan}, \citenamefont {Kuppurao}, \citenamefont
  {Lodha},\ and\ \citenamefont {Ganguly}}]{Srinivasan2012}%
  \BibitemOpen
  \bibfield  {author} {\bibinfo {author} {\bibfnamefont {V.~S.~S.}\
  \bibnamefont {Srinivasan}}, \bibinfo {author} {\bibfnamefont
  {S.}~\bibnamefont {Chopra}}, \bibinfo {author} {\bibfnamefont
  {P.}~\bibnamefont {Karkare}}, \bibinfo {author} {\bibfnamefont
  {P.}~\bibnamefont {Bafna}}, \bibinfo {author} {\bibfnamefont
  {S.}~\bibnamefont {Lashkare}}, \bibinfo {author} {\bibfnamefont
  {P.}~\bibnamefont {Kumbhare}}, \bibinfo {author} {\bibfnamefont
  {Y.}~\bibnamefont {Kim}}, \bibinfo {author} {\bibfnamefont {S.}~\bibnamefont
  {Srinivasan}}, \bibinfo {author} {\bibfnamefont {S.}~\bibnamefont
  {Kuppurao}}, \bibinfo {author} {\bibfnamefont {S.}~\bibnamefont {Lodha}}, \
  and\ \bibinfo {author} {\bibfnamefont {U.}~\bibnamefont {Ganguly}},\ }\href
  {\doibase 10.1109/LED.2012.2209394} {\bibfield  {journal} {\bibinfo
  {journal} {IEEE Electron Device Lett.}\ }\textbf {\bibinfo {volume} {33}},\
  \bibinfo {pages} {1396} (\bibinfo {year} {2012})}\BibitemShut {NoStop}%
\bibitem [{\citenamefont {Chen}\ \emph {et~al.}(2013)\citenamefont {Chen},
  \citenamefont {Komura}, \citenamefont {Degraeve}, \citenamefont {Govoreanu},
  \citenamefont {Goux}, \citenamefont {Fantini}, \citenamefont {Raghavan},
  \citenamefont {Clima}, \citenamefont {Zhang}, \citenamefont {Belmonte},
  \citenamefont {Redolfi}, \citenamefont {Kar}, \citenamefont {Groeseneken},
  \citenamefont {Wouters},\ and\ \citenamefont {Jurczak}}]{Chen2013}%
  \BibitemOpen
  \bibfield  {author} {\bibinfo {author} {\bibfnamefont {Y.~Y.}\ \bibnamefont
  {Chen}}, \bibinfo {author} {\bibfnamefont {M.}~\bibnamefont {Komura}},
  \bibinfo {author} {\bibfnamefont {R.}~\bibnamefont {Degraeve}}, \bibinfo
  {author} {\bibfnamefont {B.}~\bibnamefont {Govoreanu}}, \bibinfo {author}
  {\bibfnamefont {L.}~\bibnamefont {Goux}}, \bibinfo {author} {\bibfnamefont
  {A.}~\bibnamefont {Fantini}}, \bibinfo {author} {\bibfnamefont
  {N.}~\bibnamefont {Raghavan}}, \bibinfo {author} {\bibfnamefont
  {S.}~\bibnamefont {Clima}}, \bibinfo {author} {\bibfnamefont
  {L.}~\bibnamefont {Zhang}}, \bibinfo {author} {\bibfnamefont
  {A.}~\bibnamefont {Belmonte}}, \bibinfo {author} {\bibfnamefont
  {A.}~\bibnamefont {Redolfi}}, \bibinfo {author} {\bibfnamefont {G.~S.}\
  \bibnamefont {Kar}}, \bibinfo {author} {\bibfnamefont {G.}~\bibnamefont
  {Groeseneken}}, \bibinfo {author} {\bibfnamefont {D.~J.}\ \bibnamefont
  {Wouters}}, \ and\ \bibinfo {author} {\bibfnamefont {M.}~\bibnamefont
  {Jurczak}},\ }\href {\doibase 10.1109/IEDM.2013.6724598} {\bibfield
  {journal} {\bibinfo  {journal} {Tech. Dig. - Int. Electron Devices Meet.
  IEDM}\ ,\ \bibinfo {pages} {252}} (\bibinfo {year} {2013})}\BibitemShut
  {NoStop}%
\bibitem [{\citenamefont {Kuhn}(2008)}]{Kuhn2008}%
  \BibitemOpen
  \bibfield  {author} {\bibinfo {author} {\bibfnamefont {K.~J.}\ \bibnamefont
  {Kuhn}},\ }\href@noop {} {\bibfield  {journal} {\bibinfo  {journal} {IEDM
  2008 Present.}\ } (\bibinfo {year} {2008})}\BibitemShut {NoStop}%
\bibitem [{\citenamefont {Dimitriadis}\ \emph {et~al.}(1995)\citenamefont
  {Dimitriadis}, \citenamefont {Logothetidis},\ and\ \citenamefont
  {Alexandrou}}]{Dimitriadis1995}%
  \BibitemOpen
  \bibfield  {author} {\bibinfo {author} {\bibfnamefont {C.~A.}\ \bibnamefont
  {Dimitriadis}}, \bibinfo {author} {\bibfnamefont {S.}~\bibnamefont
  {Logothetidis}}, \ and\ \bibinfo {author} {\bibfnamefont {I.}~\bibnamefont
  {Alexandrou}},\ }\href {\doibase 10.1063/1.114070} {\bibfield  {journal}
  {\bibinfo  {journal} {Appl. Phys. Lett.}\ }\textbf {\bibinfo {volume}
  {502}},\ \bibinfo {pages} {502} (\bibinfo {year} {1995})}\BibitemShut
  {NoStop}%
\bibitem [{\citenamefont {Moravej}\ and\ \citenamefont
  {al.}(2004)}]{Moravej2004}%
  \BibitemOpen
  \bibfield  {author} {\bibinfo {author} {\bibfnamefont {M.}~\bibnamefont
  {Moravej}}\ and\ \bibinfo {author} {\bibfnamefont {E.}~\bibnamefont {al.}},\
  }\href {\doibase 10.1088/0963-0252/13/1/002} {\bibfield  {journal} {\bibinfo
  {journal} {Plasma Sources Sci. Technol.}\ }\textbf {\bibinfo {volume} {13}},\
  \bibinfo {pages} {8} (\bibinfo {year} {2004})}\BibitemShut {NoStop}%
\bibitem [{\citenamefont {Nouchi}(2014)}]{Nouchi2014}%
  \BibitemOpen
  \bibfield  {author} {\bibinfo {author} {\bibfnamefont {R.}~\bibnamefont
  {Nouchi}},\ }\href {\doibase 10.1063/1.4901467} {\bibfield  {journal}
  {\bibinfo  {journal} {J. Appl. Phys.}\ }\textbf {\bibinfo {volume} {116}}
  (\bibinfo {year} {2014}),\ 10.1063/1.4901467}\BibitemShut {NoStop}%
\bibitem [{\citenamefont {Nagano}\ \emph {et~al.}(2007)\citenamefont {Nagano},
  \citenamefont {Tsutsui}, \citenamefont {Nouchi}, \citenamefont {Kawasaki},
  \citenamefont {Ohta}, \citenamefont {Kubozono}, \citenamefont {Takahashi},\
  and\ \citenamefont {Fujiwara}}]{Nagano2007}%
  \BibitemOpen
  \bibfield  {author} {\bibinfo {author} {\bibfnamefont {T.}~\bibnamefont
  {Nagano}}, \bibinfo {author} {\bibfnamefont {M.}~\bibnamefont {Tsutsui}},
  \bibinfo {author} {\bibfnamefont {R.}~\bibnamefont {Nouchi}}, \bibinfo
  {author} {\bibfnamefont {N.}~\bibnamefont {Kawasaki}}, \bibinfo {author}
  {\bibfnamefont {Y.}~\bibnamefont {Ohta}}, \bibinfo {author} {\bibfnamefont
  {Y.}~\bibnamefont {Kubozono}}, \bibinfo {author} {\bibfnamefont
  {N.}~\bibnamefont {Takahashi}}, \ and\ \bibinfo {author} {\bibfnamefont
  {A.}~\bibnamefont {Fujiwara}},\ }\href {\doibase 10.1021/jp0708751}
  {\bibfield  {journal} {\bibinfo  {journal} {J. Phys. Chem. C}\ }\textbf
  {\bibinfo {volume} {111}},\ \bibinfo {pages} {7211} (\bibinfo {year}
  {2007})}\BibitemShut {NoStop}%
\bibitem [{\citenamefont {Bozyigit}\ \emph {et~al.}(2015)\citenamefont
  {Bozyigit}, \citenamefont {Lin}, \citenamefont {Yazdani}, \citenamefont
  {Yarema},\ and\ \citenamefont {Wood}}]{Bozyigit2015}%
  \BibitemOpen
  \bibfield  {author} {\bibinfo {author} {\bibfnamefont {D.}~\bibnamefont
  {Bozyigit}}, \bibinfo {author} {\bibfnamefont {W.~M.~M.}\ \bibnamefont
  {Lin}}, \bibinfo {author} {\bibfnamefont {N.}~\bibnamefont {Yazdani}},
  \bibinfo {author} {\bibfnamefont {O.}~\bibnamefont {Yarema}}, \ and\ \bibinfo
  {author} {\bibfnamefont {V.}~\bibnamefont {Wood}},\ }\href {\doibase
  10.1038/ncomms7180} {\bibfield  {journal} {\bibinfo  {journal} {Nat.
  Commun.}\ }\textbf {\bibinfo {volume} {6}},\ \bibinfo {pages} {6180}
  (\bibinfo {year} {2015})}\BibitemShut {NoStop}%
\bibitem [{\citenamefont {Berger}(1996)}]{Berger1996}%
  \BibitemOpen
  \bibfield  {author} {\bibinfo {author} {\bibfnamefont {P.~R.}\ \bibnamefont
  {Berger}},\ }\bibfield  {booktitle} {\emph {\bibinfo {booktitle} {IEEE
  Potentials}},\ }\href {\doibase 10.1109/45.489734} {\ \textbf {\bibinfo
  {volume} {15}},\ \bibinfo {pages} {25} (\bibinfo {year} {1996})}\BibitemShut
  {NoStop}%
\end{thebibliography}%
\end{document}